# ShAppliT: A Novel Broker-mediated Solution to Generic Application Sharing in a Cluster of Closed Operating Systems

[*1]Chen Guo, [2]Cenzhe Zhu, [3]Teng Tiow Tay

*ECE Department, National University of Singapore,*
[1]*guochen@nus.edu.sg,* [2]*cenzhe.zhu@nus.edu.sg,* [3]*eletaytt@nus.edu.sg*

***Abstract.*** **With advances in hardware and networking technologies and mass manufacturing, the cost of high end hardware had fall dramatically in recent years. However, software cost still remains high and is the dominant fraction of the overall computing budget. Application sharing is a promising solution to reduce the overall IT cost. Currently software licenses are still based on the number of copies installed. An organization can thus reduce the IT cost if the users are able to remotely access the software that is installed on certain computer servers instead of running the software on every local computer. In this paper, we propose a generic application sharing architecture for users' application sharing in a cluster of closed operating systems such as Microsoft Windows. We also propose a broker-mediated solution where we allow multiple users to access a single user software license on a time multiplex basis through a single logged in user. An application sharing tool called ShAppliT has been introduced and implemented in Microsoft Windows operating system. We evaluated their performance on CPU usage and memory consumption when a computer is hosting multiple concurrent shared application sessions.**

**Keywords***: Cluster Computing, Peer to Peer Network, Application Sharing, Remote Access, Software License, Windows Operating System*

## 1. Introduction

Application sharing is a promising solution to effectively reduce the overall cost of computing. The greatest benefit of application sharing is that software can be remotely used by the users from their local computers which may have incompatible operating system and lower processing power required by the software. This is because the users are not actually running the software on their local computer, but remotely accessing and controlling the desktop (and therefore the software) of the host computer. With the use of the application sharing software, it is possible for individuals and organization to save huge amount of money that they would have spent on purchasing more copies of software to cater for all of the local computers.

With increasing performance of general purpose computer and high speed communication, cluster computing is becoming a promising research area. A cluster environment may consist of heterogeneous operating systems including closed/proprietary operating systems and open source operating systems. A closed operating system is one where source code is not made available. Users may license the object code, but is not at liberty to modify or change. Examples of proprietary operating systems are Windows and Mac OS X. Open source operating systems allow the user to tweak and change. Examples of open source operating systems are Linux for personal computers and Android for mobile devices. In the cluster environment, we have kept proprietary operating systems in mind in our design. We design add-ons to these systems but do not modify the source code at the operating system level. For example, the client version of Windows is designed to be used by one person at a time and the terminal service also limits the number of users logged in to one at a time [1]. Two people cannot log on and access the computer system at the same time even if it includes just a physical, local-console login and a remote login. How to perform application sharing on such a proprietary operating system is an important issue to be addressed in our research.

In this paper, we propose a novel application sharing architecture for generic application sharing in a standard local area network. We provide a broker-mediated solution to extend single user software license to multiple user usage and resolve the problem of multiple users' access to proprietary operating systems. The objectives of our work are achieved through the implementation of a peer-to-peer application sharing tool called ShAppliT. ShAppliT is a middleware residing on top of the operating system. It implements a multiple-user and resource management protocol and provides a single client access to the underlying computer system. Two versions of ShAppliT have been implemented based on Microsoft Windows operating system. The first implementation, ShAppliT V1.0 achieves the research goal by modifying by Windows and the second implementation, ShAppliT V2.0 by using a broker-mediated solution to support concurrent application sharing sessions. Performance is evaluated and compared between ShAppliT V1.0 and ShAppliT V2.0 on CPU loading with multiple concurrent sessions.

This paper is organized as follows. In Section 2, we give an over of the application sharing cluster system. In Section 3, we describe the two versions of ShAppliT implementation. In Section 4, we discuss the memory performance of ShAppliT V1.0 and V2.0, licensing issues and some limitations of current system. In Section 5, we review state of the art application and desktop sharing products and communication protocols for application sharing. In the last Section, we provide concluding remarks and challenges for future work.

## 2. System overview

As shown in Figure 1, each node with ShAppliT in the cluster is called a peer. It can act as an application provider or/and as an application consumer. All the computers are connected via a high speed local area network (LAN). Computers with ShAppliT installed form a cluster network within the LAN to facilitate handshaking, message exchanging and remote desktop connections that are exclusive for ShAppliT users [2].

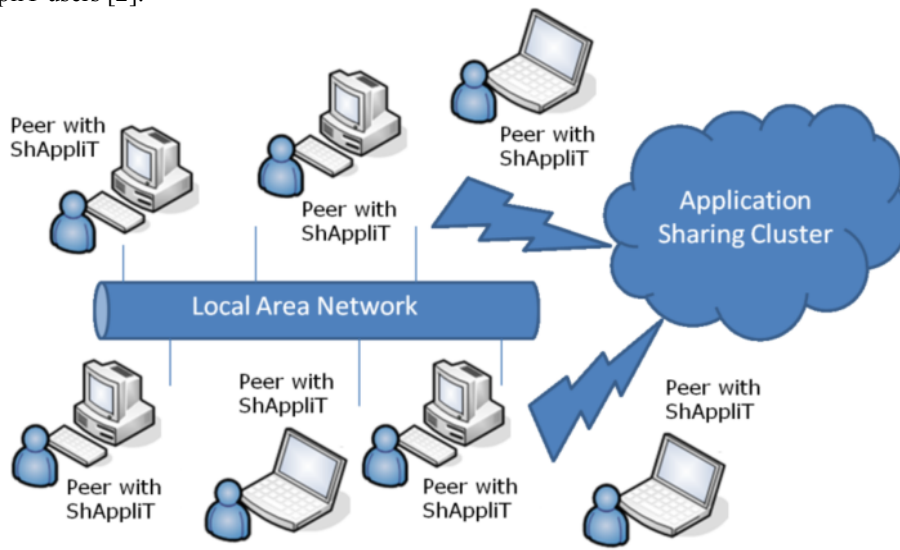

**Figure 1.** Application Sharing Cluster System overview [2]

In our current ShAppliT system, we provide a technique to coordinate multiple users' assessments for closed system using a broker-mediated mechanism. As shown in Figure 2, there is a layer on top of all operating systems for multiple user and resource management. It works as a broker to receive request from multiple users and manage the session for each user and only have one access to the operating system. The operating system, together with the underlying applications and resources fulfill the broker's requests. Our application sharing tool acts as the bridge between the clients and the server. Only one master session logs in to the application server and accesses the host Windows OS via terminal service. All the tasks are received by the broker from multiple clients, both remote and local computer users. Therefore, the server sees only one remote desktop session and does work for the broker only. The broker takes over the responsibility of negotiation with remote clients, forwards the input events to the server OS and redirects the display data back to the respective clients. In a way it shares a single-user application among multiple clients via a single log in to that application.

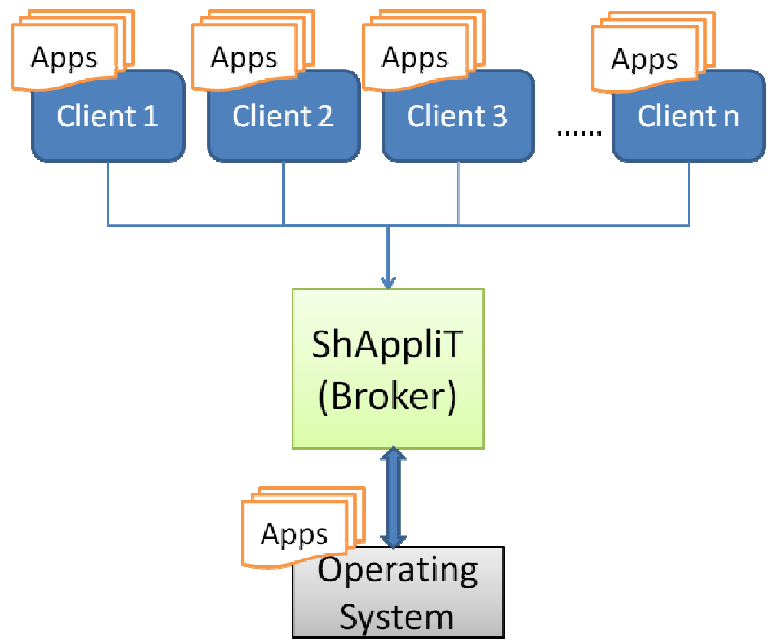

**Figure 2.** Broker-mediated Application Sharing on Closed Systems [2]

## 3. Implementation

The application, ShAppliT realizes the proposed peer-to-peer application sharing on closed systems in a cluster. It is implemented on Windows XP x86 32-bit operating system in a LAN. Unlike Linux which is a multi-user system designed to handle multiple concurrent users, Windows client systems are designed to be used by one person at a time [3]. Windows XP is typically used by standalone users whereas Window Server 2003 is normally deployed as a server operating system built to support multiple clients concurrently. However, Windows Server 2003 contains complex functionality and is mainly operated by programmers or administrators and it is many times costlier than XP, which make Windows Server 2003 not desirable for peer to peer usage. Since Windows XP is a single user operating system, it is an obstacle to the realization of peer-to-peer application sharing.

ShAppliT is divided into three parts:
- Initialization and management of a cluster
- Incoming and outgoing packet management
- Establishment of multiple remote application sharing sessions

Operation is as follows. Upon launch, the App Pool Management shown in Figure 3 searches through the system registry to track all applications installed on the host computer and generates a list. Subsequently, Sharing Permission Setting component allows the user to configure which application to be offered for sharing via the Application Pool Management. The component, Cluster Joining Management handles all the handshaking protocol in the creation/joining of a cluster.

Query Sending Management creates queries for applications. The creation of a query is stored in a list and is periodically broadcasted by Query Sending Management across the cluster network.

Simultaneously, Request Listening Management has an opened port that listens to requests broadcasted in the cluster network. All requests received will be stored in a list in Request Listening Management. Request Listening Management will periodically process the requests in the list by verifying whether relevant conditions as specified by the users are met. If all conditions are met, Request Listening Management will broadcast a message containing the information required to establish a remote desktop session across the network with the requester. Upon receiving this message, the requester will launch a Remote Desktop Connection using the information contained in the message.

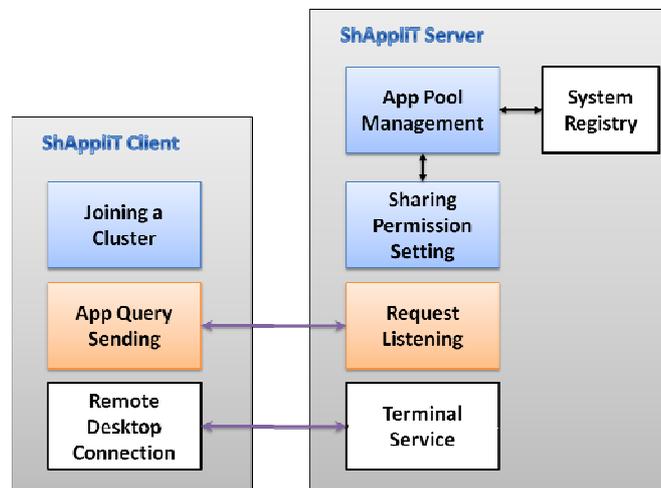

**Figure 3.** System architecture of ShAppliT V1.0

### 3.1. Cluster management

In our first attempt to create a peer-to-peer application sharing cluster, we implemented Microsoft Peer Name Resolution Protocol (PNRP) as our base protocol [4]. However, the result is not satisfactory because of an excessive delay in the connection. Therefore, we have implemented a new system using multicast approach. Multicast packet is addressed using a single identifier for a group of receivers. This address indirection allows a copy of the packet that is addressed to the group to be delivered to all the multicast receivers associated with that group.

Classful network as used by multicast is succeeded by classless inter-domain routing. However, multicasting address is still considered as Class D address. Classless inter-domain routing used significant bits to represent host and network. For an example, 192.168.0.0/16 means that there are 2^(32-16) host in the network and they start from 192.168.0.0 to 192.168.255.255. Figure 4 shows a class D identifier, 234.5.6.7, which is used to associate a group of receivers. This group is referred as a multicast group. Figure 5 describes the implementation of multicast clustering using Win32 APIs.

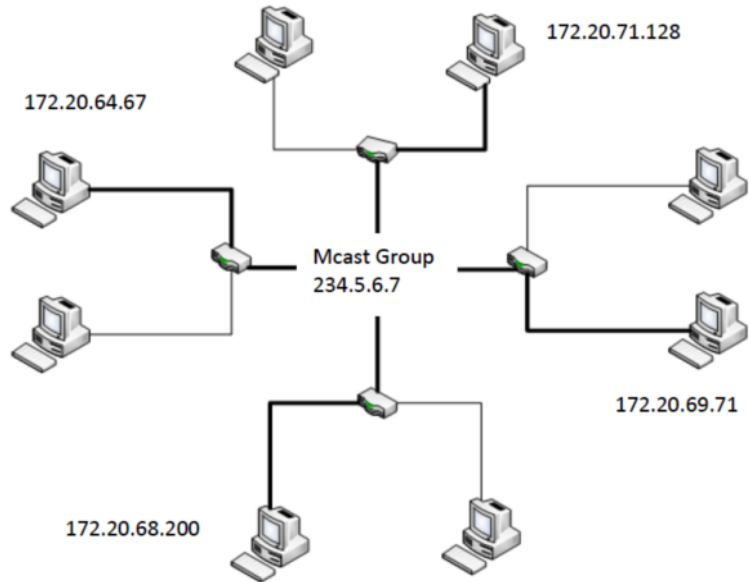

**Figure 4.** Multicast group

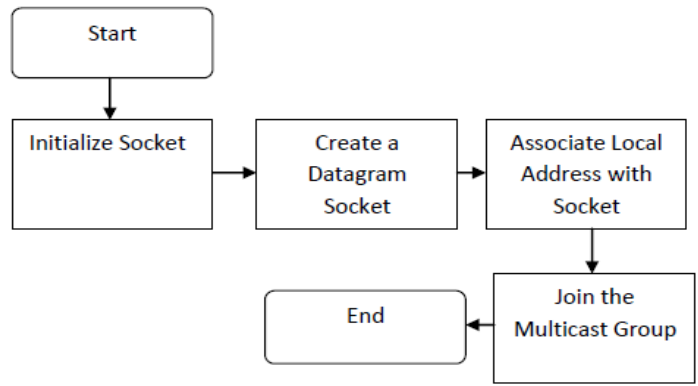

**Figure 5.** Flow chart of joining a multicast group

### 3.2. Incoming and outgoing packet management

The message passed within the cluster determines the sender, the message type and the application requested. In the example below, suppose Alice request WinWord from the multicast group. Bob replies to Alice's request. Charlie discards Alice's request because request has been fulfilled by Bob. We define four message header types:

Type 0: request an application by Alice, example: 0//winword.exe

Type 1: reply a particular request, example:1//Alice_IP//winword.exe//C:\\Program Files\\Microsoft Office\\winword.exe//guest2

Type 2: handshake between all hosts to notify each other their existence in the cluster

Type 3: graceful disconnection if a host is to leave the cluster

When ShAppliT receive datagram from the network, these packets are stored in the list. There are 3 kind of list:

A list of all requests by the host
A list of all incoming replies to the request of the host
A list of all incoming requests from other clients

The information in the lists must be unique. This uniqueness can be enforced by using STL (Standard Template Library) set [5]. Sets are associate containers that store unique elements or keys. The uniqueness of the structure is enforced by the operator of the structure. A thread is used to process incoming datagram stored in the set. The methods used are described in Figure 6.

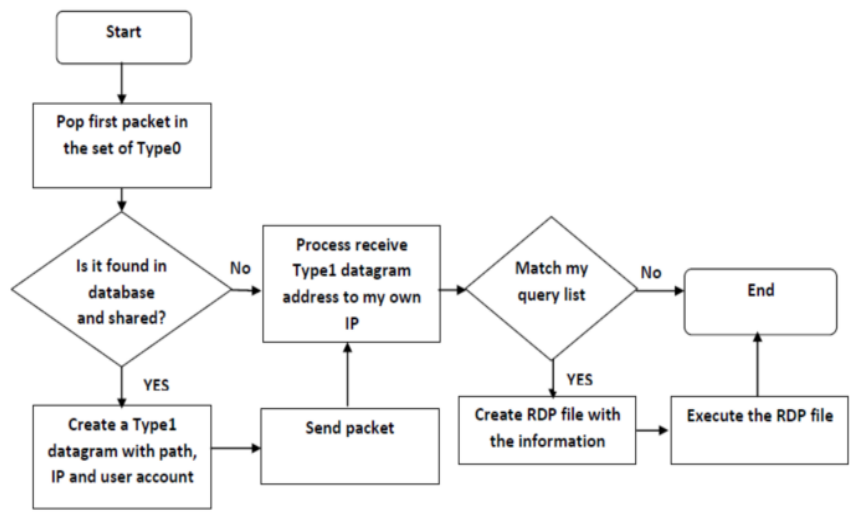

**Figure 6.** Flow chart for processing datagram in set container

QueryPacket structure: there is no duplicate application name. Example: Alice cannot request winword.exe twice until the previous request is timed out or is satisfied by other peer in the cluster.

RecReplyPac structure: this structure is used to store replies of all the requests made. The uniqueness of the structure is enforced by "a peer does not satisfy any request twice". Example: Alice will not store the reply packet on winword.exe from Bob twice.

KeyValueRec structure: this structure is used to process incoming request packets. It stores a temporary list for later processing. This structure's uniqueness is enforced by "the same IP should not associate to the same application". Example: Bob receives Alice request on winword.exe and powerpnt.exe, Bob should not receive Alice's request on winword.exe twice.

Figure 7 shows C++ codes for message structures used to store the receiving packets from the cluster.

```
typedef struct _RecReplyPac{
wstring strAppName;
wstring strIpv4;
wstring strFullPathName;
wstring strUsername;
bool operator<(const _RecReplyPac& A) const
{
return (strIpv4.compare(A.strIpv4) < 0 &&
strAppName.compare(A.strAppName) < 0 );
}
```

```
}RecReplyPac;
typedef struct _QueryPacket{
SYSTEMTIME systemTime;
wstring strAppName;
bool operator<(const _QueryPacket& A) const
{
return (strAppName.compare(A.strAppName) < 0 );
}
}QueryPacket;
typedef struct _KeyValueRec{
wstring strIpv4;
wstring strAppName;
bool operator<(const _KeyValueRec& A) const
{
return (strIpv4.compare(A.strIpv4) <0) ^
(strAppName.compare(A.strAppName) < 0);
}
}KeyValueRec;
```

**Figure 7.** C++ codes of message structures used to store the receiving packet from the cluster

### 3.3. Establish remote application sessions

#### 3.3.1. ShAppliT V1.0

ShAppliT V1.0 supports remote application sessions concurrently by making modifications on terminal service (TS) DLL file and the registry of Windows XP as described in references [6] and [7]. The changes may violate the license stated by Microsoft. Our main goal is to demonstrate the feasibility and research on the future development of peer-to-peer clustering concept. In this case, Windows terminal service server manages the connections sessions directly such that no broker is needed for exchange of information between server and client.

#### 3.3.2. ShAppliT V2.0

With ShAppliT V2.0 using the broker-mediated method of establishing multiple remote application sessions, we provide a broker-mediated solution to extend single user software license for multiple-user usage and solve the problem of working on closed or proprietary operating systems.

Instead of managing multiple connections using Windows terminal service server, we implement ShAppliT V2.0 Server which sits in between the ShAppliT V2.0 client and Windows TS server as a broker. It handles tasks from multiple clients and passes them to the TS server. Therefore, TS server sees only one Remote Desktop Protocol (RDP) session and does work for the ShAppliT Server only. And ShAppliT Server takes over the responsibility of negotiation with remote clients, forwards the input events to TS server and redirects the display data back to the respective clients. Figure 8 shows the detailed programming model of ShAppliT V2.0.

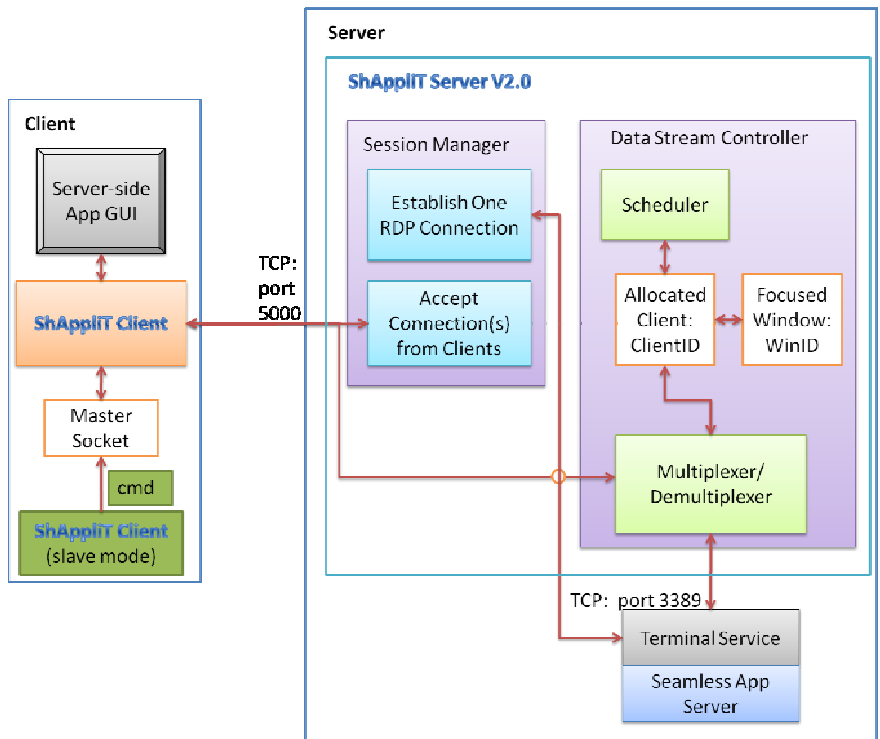

**Figure 8.** Programming model of ShAppliT V2.0 system

Master mode is the default mode of ShAppliT Client. When run in master mode, ShAppliT Client creates and listens on a master socket. After creation of a remote user session with a ShAppliT Server and maintenance of that connection, ShAppliT Client listens on the master socket and checks master socket each time when TCP layer receives packets.

When run in slave mode, ShAppliT Client notifies the master Client instance of a new command to be run by sending command (e.g. "mspaint") to the master socket and then exits. The master instance detects a command from a client and sends a client-to-server message (e.g. "spawn, mspaint") to the ShAppliT Server. The message will be directed to SeamlessApp server component at the Windows server, which runs the new command on the server machine. Finally, a remote application is launched at the Windows server and the application Graphic User Interface (GUI) will be received by ShAppliT Client. Moreover, we can use the slave mode multiple times to send more application commands. So, it provides connection sharing by allowing a single ShAppliT connection to launch multiple applications.

There are two components in the ShAppliT Server V2.0, namely the Session Manager and Data Stream Controller. The session manager component first establishes an RDP connection with Microsoft Terminal Service Server. Then, it listens on TCP port 5000 and accepts connections from remote clients. It creates new user sessions for remote clients after successful connection negotiation. The connection sequence follows the RDP connection sequence mentioned in MS-RDPBCGR [8]. After a new user session is created, a new client is added to the data stream controller for starting additional application. Data stream controller is in charge of multiplexing and de-multiplexing the clients' and server's traffic. It maintains the smooth execution of multiple remote user sessions of ShAppliT V2.0 system.

There are three major programming modules inside data stream controller, namely the scheduler, MUX and DEMUX. There are two important control signals: Allocated Client: Client ID and Focused Window: Win ID. Each client has at most one focused window. Data stream controller keeps track of the focused window for each client. According to the allocated client information determined by the scheduler, data stream controller sends over the focus window information to TS Server then followed by the client's input events. Our current scheduler uses a round-robin scheduling algorithm. Clients' input events are queued in a buffer of each client respectively. The scheduler assigns time slices to each client in equal portions and in circular order. The next client will be allocated after the timer expired. Allocated client is the control signal for the steam multiplexer and de-multiplexer. At a time only one client is enabled to transmit its input events and to receive the graphic updates from server. The allocated client is determined by a scheduler. Clients' input events in the global channel including the keyboard and mouse inputs are buffered in an event queue of each client respectively.

    Figure 9 illustrates the relationship of focused window and allocated client at both client and server sides. Each client may have one or more applications running from the same sever, but at a time there is only one window focused by the client. A focused window is the window the client is operating on currently. So, at the client side each client will have at most one window focused shown with filled colour box. At the sever side, only one window is focused each time shown with filled colour box. Therefore, it is important to keep track of the focused window ID for each client. Allocated client is decided by the scheduler according to the scheduling algorithm as mentioned earlier. When it comes to a client's turn to send over its events the server will be notified about the current focused window by our ShAppliT Server. Then the TS server will perform operations on the focused window of the allocated client according to the input events received and send the server output graphic update data over to the allocated client.

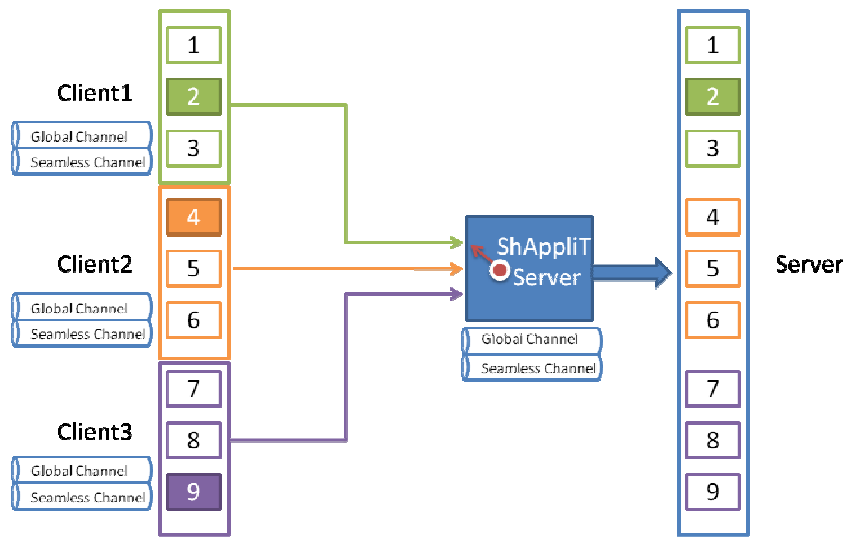

**Figure 9.** Illustration of focused window and allocated client

    Focused window information is extracted from the network packet flow of the seamless virtual channel in RDP as mentioned in the previous section. The seamless channel ID is determined by the negotiation between client and server during the RDP connection sequence. Focused window information is carried in the seamless virtualChannelData field with the format "focus, win ID, flags". The seamless channel data is directed by the TS server to the SeamlessApp Server endpoint for further process.

## 4. Results and discussion

### 4.1. Multi-session load analysis

   Our test bed is set up in a LAN consisting of ten computers with ShAppliT and identical system environment configurations. All PCs are Pentium 4 3.0GHz machines with 512MB physical memory running Windows XP professional SP3. The performance evaluation is mainly focused on describing the impact of increasing the number of remote application sessions on the memory consumption of a host computer running ShAppliT. The load analysis shows that additional remote connection results in a linear increase of the commit charges on host computer. A comparison of memory performance is made between ShAppliT V1.0 and ShAppliT V2.0. ShAppliT V2.0 consumes less commit charges on host computer. Furthermore, lawful issue of sharing licensed applications is solved in ShAppliT V2.0. We will further discuss this in session 4.2.

   In the experiment, we test on a single host machine running multiple remote application sessions of WordPad.exe. This analysis helps us evaluate the memory performance of the computer and determine the maximum concurrent session to be accepted on the machine. This testing is conducted using a host computer with ShAppliT V1.0 or ShAppliT V2.0 installed and deployed. The performance data is obtained using the performance analysis tool implemented in Windows Task Manager. Table 1 below gives an overview on detailed performance analysis of Windows Task Manager.

Table 1. Details on Windows Task Manager Performance analysis [9]

| Parameter | Details |
| --- | --- |
| Commit Charge | Amount of virtual memory reserved by the operating system for the process. Memory allocated to programs and the operating system. Because of memory copied to the paging file, called virtual memory, the value listed under Peak may exceed the maximum physical memory. The value for Total is the same as that depicted in the Page File Usage History graph. |
| Physical Memory | The total physical memory, also called RAM, installed on your computer. Available represents the amount of free memory that is available for use. The System Cache shows the current physical memory used to map pages of open files. |
| Kernel Memory | Memory used by the operating system kernel and device drivers. The paged is memory that can be copied to the paging file, thereby freeing the physical memory. The physical memory can then be used by the operating system. Non-paged is memory that remains resident in physical memory and will not be copied out to the paging file. |

   We have used a control session as a reference whereby the data is captured when no remote session is taking place. The data is being recorded every time an additional remote session is launched from a client computer and the result is shown in Table 2 and Table 3.

Table 2. Multi-session load analysis on host computer with ShAppliT V1.0

| No. of remote sessions | Total physical memory(KB) | Available physical memory(KB) | Total kernel memory(KB) | Paged kernel memory(KB) | Total commit charge(KB) |
|---|---|---|---|---|---|
| 0 | 514116 | 318056 | 37864 | 27276 | 248176 |
| 1 | 514116 | 278588 | 42312 | 31420 | 268272 |
| 2 | 514116 | 269660 | 44660 | 33600 | 285812 |
| 3 | 514116 | 260704 | 47076 | 35876 | 300284 |
| 4 | 514116 | 252144 | 49412 | 38056 | 314888 |
| 5 | 514116 | 261168 | 51440 | 40036 | 318972 |
| 6 | 514116 | 252172 | 53752 | 42188 | 343584 |
| 7 | 514116 | 236348 | 55448 | 43768 | 358156 |
| 8 | 514116 | 227700 | 57600 | 45776 | 372684 |
| 9 | 514116 | 220752 | 59864 | 47896 | 388188 |

Table 3. Multi-session load analysis on host computer with ShAppliT V2.0

| No. of remote sessions | Total physical memory(KB) | Available physical memory(KB) | Total kernel memory(KB) | Paged kernel memory(KB) | Total commit charge(KB) |
|---|---|---|---|---|---|
| 0 | 514116 | 318056 | 37864 | 27276 | 248176 |
| 1 | 514116 | 308092 | 38112 | 27524 | 249012 |
| 2 | 514116 | 306060 | 38244 | 27656 | 251132 |
| 3 | 514116 | 305416 | 38388 | 27800 | 251932 |
| 4 | 514116 | 303880 | 38532 | 27944 | 254080 |
| 5 | 514116 | 302248 | 38712 | 28124 | 255744 |
| 6 | 514116 | 300976 | 38868 | 28280 | 257356 |
| 7 | 514116 | 308312 | 39020 | 28432 | 258932 |
| 8 | 514116 | 308716 | 39172 | 28584 | 260468 |
| 9 | 514116 | 307864 | 39324 | 28736 | 261856 |

The load analysis of ShAppliT V1.0 (Figure 10) shows that additional remote connection results is a linear increase of the commit charge on the host computer. And the physical memory at the host computer decreases with increasing number of multiple remote sessions. As such, it is necessary to set a limit on the maximum number of concurrent sessions so that the host computer would not be burdened by excessive remote connections and experience laggings in the local session. This result also highlights that although this system provides certain degree of scalability but further performance optimization on memory consumption still need to be done.

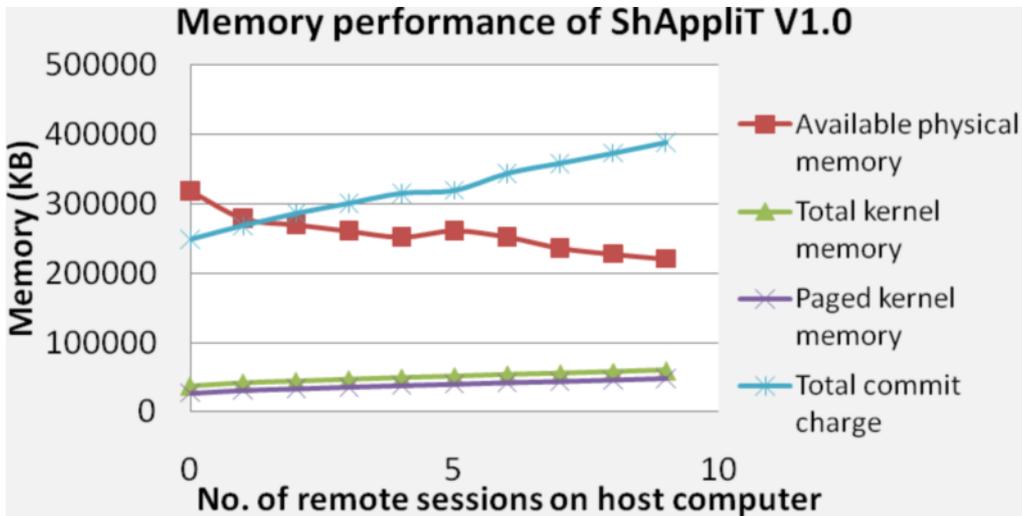

Figure 10. Memory performance of ShAppliT V1.0 when hosting multiple remote sessions

The load analysis of ShAppliT V2.0 (Figure 11) shows that additional remote connection results in a linear increase of the commit charge on host computer. The increment of commit charge is very small with increasing number of concurrent remote sessions. In addition, the available physical memory of the host computer is affected very little by the multiple remote sessions.

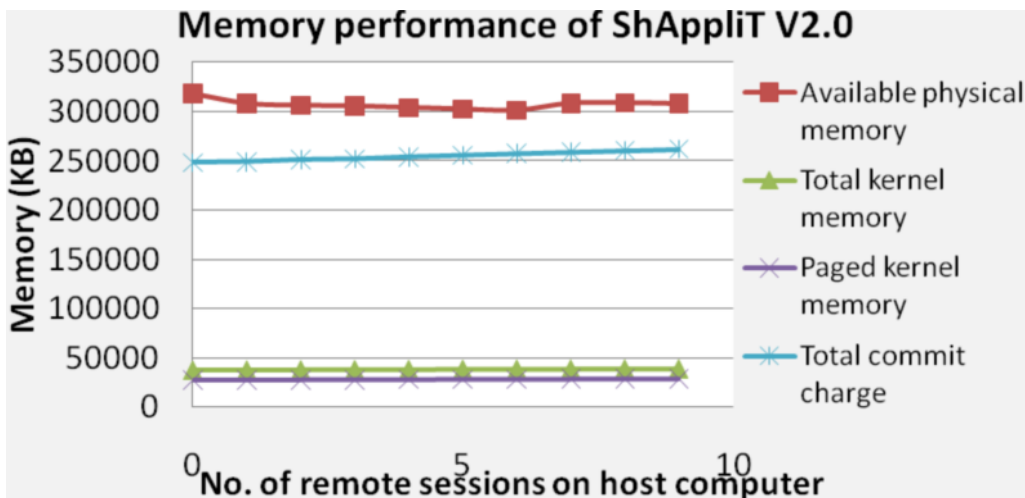

Figure 11. Memory performance of ShAppliT V2.0 when hosting multiple remote sessions [2]

Memory management is more effective in ShAppliT V2.0 compared to ShAppliT V1.0 observed from Figure 12 below. This is because in ShAppliT V2.0 there is only one RDP connection established and maintained by ShAppliT Server. Each additional application launched at host computer is invoked by SeamlessApp server in the same way as using cmd.exe at host computer. As a result, starting a new application session, the operating system only allocates the necessary memory resource to the application process running within the same user. While, in ShAppliT V1.0 multiple remote connections are made to Windows TS server directly and multiple RDP sessions are established. Each time any new request of application from client, the host computer launches an additional RDP session for the application. Therefore, the operating system reserves the memory for multiple RDP sessions in ShAppliT V1.0, which consumes much more memory than within one RDP connection session.

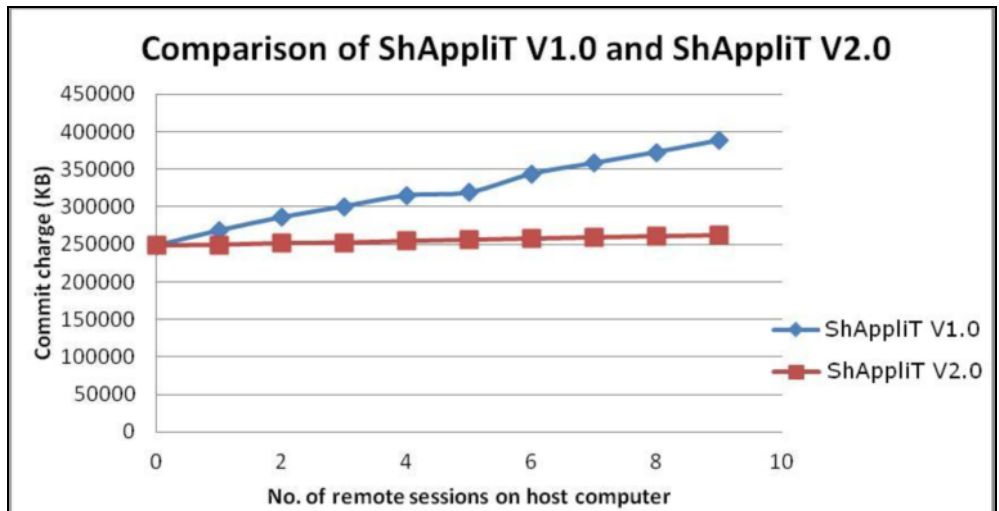

**Figure 12**. Comparison between ShAppliT V1.0 and ShAppliT V2.0 on commit charge when hosting multiple remote sessions

### 4.2. License issue on application sharing

In our ShAppliT V1.0 system there is a legal issue associated with the software license. Most of the software installed in personal computers have a single user software license and cannot be transferred from one user to another. For example Microsoft office edition 2007 says the single primary user of a licensed device may access and use the software installed on the licensed device. Single user may use remote access technologies, such as the Remote Desktop features in Microsoft Windows or NetMeeting, to access and use the licensed copy of the Software, provided that only the primary user of the device hosting the remote desktop session accesses and uses the Software with a remote access device [9].

The single user licensing problem of application sharing is solved in our current approach ShAppliT V2.0 by establishing one RDP connection for multiple clients. The ShAppliT Server sits in between the ShAppliT client and TS server as a broker. It logs in to the host operating system via RDP, handles tasks from multiple clients, including the local user sitting in front of the computer and passes them to the TS server. Therefore, TS server sees only one RDP session and it feels that it works for the broker only. And the broker is in charge of connecting to TS Server, creating remote user sessions and multiplexing/de-multiplexing the data streams. So, when a client want to launch a remote application from the server, the application will be open at the server under the same user account which is the one established by the broker. Therefore, with our application sharing tool ShAppliT, as long as there is one user license for the software, it can be shared among multiple clients without violating any licensing terms.

Furthermore, the client version of Microsoft Windows operating system (e.g. Windows XP Professional, Windows 7) terminal service limits the number of users logged in to one at a time. Two people cannot be logged on at the same time even if it includes just a physical, local-console login and a remote login. It has to be one or the other and only one user at a time. In ShAppliT V1.0, we modify the DLL of terminal service to allow multiple users logging in. The changes may violate the license stated by Microsoft. In ShAppliT V2.0 system, we only have one master user login by ShAppliT Server (broker). So, the problem of the closed system limitation on single user logged-in session is not an issue for broker mediated application sharing system.

### 4.3. Some limitations of ShAppliT V2.0

In our current implementation of application sharing cluster, we choose Windows OS as our implementation platform. However, in a cluster environment the operating systems are heterogeneous in general. More implementation and performance testing should be done on other closed systems as well, for example Mac OS. We also assume all users are honest and trustworthy and therefore security management is not considered at user level in our current design. Furthermore, reliable service is not guaranteed as peers can fail or shut down at any time. Fault recovery mechanism will be done in our future work. Finally, resource management needs to be improved. Finally, requested resources are allocated to a user when the first response is received from the cluster. Matchmaking and load balancing need to be applied for better resource utilization.

## 5. Related works

### 5.1. Application and desktop sharing products

Application and desktop sharing enables remote administration, group collaboration, remote trouble shooting, e-learning, and software tutoring and so on [11]. In the market, many remote control and desktop sharing solutions are available. The application sharing products use similar technology and concept to implement. Microsoft has Windows Meeting Space for Windows Vista and Netmeeting for Windows XP. Netmeeting was released in 1999 for Windows 98; in our tests it fails to display pop-ups and menus. Windows Vista introduces an application sharing feature as part of Windows Meeting Space, but all the attendees must use Windows Vista.VNC [12] is a cross-platform open source desktop sharing system but it supports only screen sharing. VNC supports multiple users but it lacks a floor control protocol. VNC uses a client-pull based transmission mechanism which performs poorly compared with server-push based transmissions under high round-trip time (RTT). SharedAppVnc [13] supports true application sharing, but the delay is on the order of seconds. It uses a loss codec and does not support multicast. TeleTeachingTool [14] and MAST [15] use multicast in order to build a scalable sharing system. TeleTeachingTool is developed just for online teaching so it does not allow participants to use the shared desktop. Also, it does not support real application sharing. MAST allows remote users to participate via their keyboard and mouse but its screen capture model is based on polling the screen which is very primitive and not comparable to current state of art the capturing methods like mirror drivers. BASS [11] is an application and desktop sharing platform which allows two or more people to collaborate on a single document, drawing or project in real-time. But when one user manipulates the application via keyboard and mouse events, other users receive the screen updates simultaneously. It does not support real application sharing.

### 5.2. Communication protocols for application sharing

There are many communication protocols defined by different vendors or organizations, such as RFB (Remote Frame Buffer) for VNC, RDP for Windows Terminal Service, and ITU-T T.128 [16] for NetMeeting and SunForum [17]. In general, most remote desktop protocols are similar in term of the functionality they offer. However, these protocols can be differentiated by the way the implementation of system layer where all the redirection of graphical output and user input take place. This is also the key component that determines the speed and quality of a remote desktop protocol. Some protocols compress the graphical images for transmission while other uses kernel level driver for transmission.

There were two ways to implement application sharing systems. The difference is the transmission of screen contents or drawing commands [17]. In the following section, we will study VNC protocol and RDP protocol to identify the key differences that separate them.

Virtual Network Computing (VPN) was originally developed by at the Olivetti Research Laboratory in Cambridge, United Kingdom [18]. It is a graphical sharing system that uses RFB protocols. Many VNC source code available nowadays are open sources under the GNU General Public License. The most popular implementations of VNC available in the market are RealVNC and UltraVNC. In shorts, RFB is a simple protocol for remote access to graphical user interface that function at the frame buffer level [19]. Therefore, it is highly versatile and applicable to applications and systems across different platforms and operating systems. As for the display side of the protocol, a low level primitive graphics concept has been applied. The data containing the graphical display information at the pixel level such as coordinate and image block of a particular group of pixels are compressed and transmitted regularly from the server to the client. In another words, the update of a display screen consists of a series of frame buffer updates that refresh the display screen block by block. The way this concept works is similar to how video frames refresh.

On the other hand, RDP is an extension of the ITU-T.128 application sharing protocol developed by Microsoft [20]. Basic connectivity and graphics remoting is designed to facilitate user interaction with a remote computer system by transferring graphics display information from the remote computer to the user and transporting input commands from the user to the remote computer, where the input commands are replayed on the remote computer. RDP also provides an extensible transport mechanism which allows specialized communication to take place between components on the user computer and components running on the remote computer. This proprietary protocol provides a mean to access the graphical interface of a remote host computer. Similar to other remote desktop applications, the processing of a running application is being done in the host computer, only the graphical presentation of the desktop is being transmitted to the client. However, as compared to VNC, RDP provides a faster remote access speed [21]. This is due to the fact that RDP hooks deeper into Windows API to optimize the information required by the client to construct the display screen. For example, while VNC is transmitting blocks of bitmap for the client to construct a display screen of a text document, RDP transmits the texts in the document itself for the client to render a display screen.

## 6. Conclusion and future work

We have developed a novel P2P application sharing system ShAppliT in a cluster which supports generic application sharing and concurrent multiple sharing sessions. The proposed architecture is a clever blend of cluster computing and peer-to-peer concepts. ShAppliT enables client remote access application resources that are not installed on the local computer. Also, a peer can host multiple remotely access sessions without any interference for his own experience. We provide a broker-mediated solution to extend a single user licensed software resource for multiple user usage without modifying the operating system in ShAppliT V2.0. Experiments also show that our application sharing system has good usability, scalability and a friendly user interface. Our broker-mediated system architecture has wide applicability on other closed systems.

Future research works are able to be carried out on security management, reliability and resource management for P2P application sharing in a cluster environment. User identification, data encryption algorithms and incentive mechanisms are ways to prevent free-riding and promote cooperation across distrustful peers [22]. In addition, a successful application sharing system should also provide reliable services. Peers can build up coordinated checkpoints [23] for fault recovery and establish redundant links across the peers in case of network failures. So, process at the failed peer can be migrated to a peer with redundant resource for fault tolerance. Resource management plays a critical rule in P2P application sharing. The research problem for resource discovery is matchmaking [24] that locates resources subject to certain constraints. Load balancing can be applied for better resource utilization [25]. Experiments will be done to evaluate the local user latency. Local user should experience fixed low latency when sharing with multiple remote sessions. When the number of remote sessions increases, the latency should at the beginning increases slowly and then keep at a certain constant acceptable value or fixed value. The value need to be calculated using a scheduling algorithm [26] and adjusted by the scheduler.

## 7. References

[1] Technical Overview of Windows Server 2003 Terminal Services, Microsoft Corporation. Retrieved 4 Jun 2012, from http://www.microsoft.com/windowsserver2003/techinfo/overview/termserv.mspx

[2] Chen Guo, Cenzhe Zhu and Teng Tiow Tay, "Sharing of Generic Single-user Application without Interference in a Cluster", The 2012 International Conference on Software and Computer Applications (ICSCA 2012), IACSIT Press, Jun 2012 Singapore.

[3] Chapter 1: Functional Comparison of UNIX and Windows. Retrieved 10 May, 2012, from http://technet.microsoft.com/en-us/library/bb496993.aspx

[4] Kurose, K. W. Computer Networking: A Top-down Approach. Pearson.

[5] Cplusplus.com. Retrieved 4 Jun 2012, from http://www.cplusplus.com/reference/stl/set/

[6] Enabling Multiple Remote Desktop Sessions in Windows XP, fawzi.wordpress.com/2008/02/09/enabling-multiple-remote-desktop-sessions-in-windows-xp/

[7] How to Enable Multiple Remote Desktop Sessions on XP or Vista, remotedesktoprdp.com/Multiple-Remote-Desktop-Sessions.aspx

[8] MSDN. MS-RDPBCGR. Retrieved 4 Jun 2012, from Remote Desktop Protocol: Basic Connectivity and Graphics Remoting Specification: http://msdn.microsoft.com/en-us/library/cc240445(v=prot.10).aspx

[9] Task Manager Overview. Retrieved 4 Jun 2012, from Windows XP Professional Product Documentation: http://www.microsoft.com/resources/documentation/windows/xp/all/proddocs/en-us/taskman_whats_there_w.mspx?mfr=true

[10] License Terms. Retrieved 4 Jun 2012, from Microsoft: www.microsoft.com/About/Legal/EN/US/IntellectualProperty/UseTerms/Default.aspx

[11] Omer, B., Henning, S. "BASS Application Sharing System", In: 10th IEEE International Symposium on Multimedia, 2008.

[12] T. Richardson, Q. Stafford-Fraser, K. R. Wood, and A. Hopper, "Virtual network computing", IEEE Internet Computing, 1998.

[13] G. Wallace and K. Li, "Virtually shared displays and user input devices", In USENIX Annual Technical Conference, 2007.

[14] P. Ziewer and H. Seidl, "Transparent TeleTeaching", In ASCILITE, pages 749–758. Auckland, New Zealand, 2002.

[15] G. Lewis, S. M. Hasan, V. N. Alexandrov, and M. T. Dove, "Facilitating collaboration and application sharing with MAST and the access grid development infrastructures", In E-SCIENCE, 2006.

[16] ITU-T T.128, Retrieved 4 Jun 2012, from http://www.itu.int/rec/recommendation.asp

[17] Hui-Chieh, L., Yen-Ping, C., Ruey-Kai, S., & Win-Tsung, L., "A Generic Application Sharing Architecture Based on Message-Oriented Middleware Platform", In Proceedings of the 10th International Conference on Computer Supported Cooperative Work in Design. IEEE, 2006.

[18] The VNC family of Remote Control Applications. Retrieved 4 June, 2012, from http://ipinfo.info/html/vnc_remote_control.php

[19] Richardson, T, the RFB Protocol, RealVNC Ltd, 2009.

[20] MSDN. The T.120 Standard. Retrieved 4 June 2012, from http://msdn.microsoft.com/en-us/library/ms709084(VS.85).aspx

[21] MSDN. Remote Desktop Protocol. Retrieved 4 Jun 2012, from http://msdn.microsoft.com/en-us/library/cc240446(v=PROT.10).aspx

[22] Han Zhao, Xinxin Liu, Xiaolin Li, "A taxonomy of peer-to-peer desktop grid paradigms", Cluster Computing, v.14 n.2, p.129-144, June 2011.
[23] Ni, L., Harwood, A., & Stuckey, P., "Realizing the e-science desktop peer using a peer-to-peer distributed virtual machine middleware", In: Proceedings of the 4th international workshop on Middleware for grid computing (MCG '06), USA, 2006.
[24] Kim, J.-S., Nam, B., Marsh, M., Keleher, P., Bhattacharjee, B., Richardson, D., Wellnitz, D., Sussman, A., "Creating a robust Desktop Grid using peer-to-peer services", In: Proc. of the IEEE International Parallel and Distributed Processing Symposium (IPDPS'07), pp. 1-7, Long Beach, CA, USA, 2007.
[25] Farrukh Arslan, "Service Oriented Paradigm for Massive Multiplayer Online Games", International Journal of Soft Computing and Software Engineering [JSCSE], Vol. 2, No. 5, pp. 35-47, 2012, Doi: 10.7321/jscse.v2.n5.4
[26] Yongbo Jiang, Zhiliang Qiu, Jian Zhang, Jun Li, "Integration of Unicast and Multicast Scheduling in Input-Queued Packet Switches with High Scalability", International Journal of Soft Computing and Software Engineering [JSCSE], Vol. 2, No. 4, pp. 14-34, 2012, Doi: 10.7321/jscse.v2.n4.2